\documentclass[aps,prl,twocolumn,unsortedaddress,superscriptaddress,floatfix]{revtex4-1}

\usepackage{amsmath,amssymb,amsbsy}
\usepackage{enumerate}
\usepackage{isotope}
\usepackage{bbm}
\usepackage{hyperref}
\usepackage{graphicx}
\usepackage{multirow,dcolumn}
\usepackage[capitalize]{cleveref}
\hypersetup{colorlinks=true,citecolor=[rgb]{0,0,0.8},
linkcolor=[rgb]{0,0,0.8},urlcolor=[rgb]{0,0,0.8}}

\usepackage[usenames]{color}
\usepackage[normalem]{ulem}

\newcommand{\ket}[1]{\left|#1\right\rangle}

\newcommand{\ev}[1]{\left\langle#1\right\rangle}

\renewcommand{\vec}[1]{\mathbf{#1}}
\newcommand{\gvec}[1]{\boldsymbol{#1}}

\newcommand{\nn}{NN}
\newcommand{\nnn}{3N}
\newcommand{\ntwolo}{N$^2$LO}
\newcommand{\nalpha}{$n$-$\alpha$}

\newcommand{\tdott}[2]{\gvec{\tau}_{#1}\!\cdot\!\gvec{\tau}_{#2}}
\newcommand{\sdots}[2]{\gvec{\sigma}_{#1}\!\cdot\!\gvec{\sigma}_{#2}}
\newcommand{\drtn}[1]{\delta_{R_{\nnn{}}}\!(#1)}

\begin{document}
\unitlength=1mm
\title{Chiral Three-Nucleon Interactions in
Light Nuclei, Neutron-$\alpha$ Scattering, \\
and Neutron Matter}

\author{J.~E.~Lynn}
\email[E-mail:~]{joel.lynn@gmail.com}
\affiliation{Theoretical Division, Los Alamos National Laboratory,
Los Alamos, New Mexico 87545, USA}
\author{I.~Tews}
\affiliation{Institut f\"ur Kernphysik,
Technische Universit\"at Darmstadt, 64289 Darmstadt, Germany}
\affiliation{ExtreMe Matter Institute EMMI, GSI Helmholtzzentrum f\"ur
Schwerionenforschung GmbH, 64291 Darmstadt, Germany}
\author{J.~Carlson}
\affiliation{Theoretical Division, Los Alamos National Laboratory,
Los Alamos, New Mexico 87545, USA}
\author{S.~Gandolfi}
\affiliation{Theoretical Division, Los Alamos National Laboratory,
Los Alamos, New Mexico 87545, USA}
\author{A.~Gezerlis}
\affiliation{Department of Physics, University of Guelph,
Guelph, Ontario, N1G 2W1, Canada}
\author{K.~E.~Schmidt}
\affiliation{Department of Physics, Arizona State University, Tempe,
Arizona 85287, USA}
\author{A.~Schwenk}
\affiliation{Institut f\"ur Kernphysik,
Technische Universit\"at Darmstadt, 64289 Darmstadt, Germany}
\affiliation{ExtreMe Matter Institute EMMI, GSI Helmholtzzentrum f\"ur
Schwerionenforschung GmbH, 64291 Darmstadt, Germany}

\begin{abstract} We present quantum Monte Carlo calculations of light
nuclei, neutron-$\alpha$ scattering, and neutron matter using local two-
and three-nucleon (\nnn{}) interactions derived from chiral effective
field theory up to next-to-next-to-leading order (\ntwolo{}).
The two undetermined \nnn{} low-energy couplings are fit to the
\isotope[4]{He} binding energy and, for the first time, to the
spin-orbit splitting in the neutron-$\alpha$ $P$-wave phase shifts.
Furthermore, we investigate different choices of local \nnn{}-operator
structures and find that chiral interactions at \ntwolo{} are able to
simultaneously reproduce the properties of $A=3,4,5$ systems and of
neutron matter, in contrast to commonly used phenomenological \nnn{}
interactions.
\end{abstract}
\pacs{21.60.--n, 21.10.--k, 21.30.--x, 21.60.De}
\maketitle

Three-nucleon (\nnn{}) interactions are essential for a reliable
prediction of the properties of light nuclei and nucleonic
matter~\cite{kalantar2011,hammer2013,carlson2014,hebeler2015,
gandolfi2015}.
In quantum Monte Carlo (QMC) calculations phenomenological \nnn{}
interactions such as the Urbana~\cite{pudliner1997} and
Illinois~\cite{pieper2001.2} models have been used with great
success~\cite{pieper2008,carlson2014}.
However, such models suffer from certain disadvantages: They are not
based on a systematic expansion and it was found that the Illinois
forces tend to overbind neutron matter~\cite{Sarsa:2003,Maris:2013}.
It is therefore unlikely that these phenomenological models can be used
to correctly predict the properties of heavy neutron-rich nuclei.

An approach which addresses these shortcomings is chiral effective field
theory (EFT)~\cite{weinberg1990,weinberg1991,epelbaum2009,machleidt2011,
hammer2013}.
Chiral EFT is a low-energy effective theory consistent with the
symmetries of quantum chromodynamics and provides a systematic expansion
for nuclear forces.
It includes contributions from long-range pion-exchange interactions
explicitly and expands the short-distance interactions into a systematic
set of contact operators accompanied by low-energy couplings fit to
experimental data.
Chiral EFT enables the determination of theoretical uncertainties and
systematic order-by-order improvement; for recent work see
Refs.~\cite{epelbaum2015,ekstrom2015,carlsson2015,furnstahl2015}.

Chiral EFT also predicts consistent many-body interactions.
In Weinberg power counting, \nnn{} forces first enter at
next-to-next-to-leading order
(\ntwolo{})~\cite{vankolck1994,epelbaum2002} and contain three
contributions: A two-pion-exchange interaction $V_C$, a
one-pion-exchange-contact interaction $V_D$, and a \nnn{} contact
interaction $V_E$.
While the first is accompanied by the couplings $c_i$ from the
pion-nucleon sector, the latter two are accompanied by the couplings
$c_D$ and $c_E$, which have to be determined in $A>2$ systems.

In addition to systematic nuclear forces, reliable many-body methods are
required to describe properties of light nuclei and of dense neutron matter.
QMC approaches, which solve the many-body Schr\"odinger equation
stochastically, are such a class of methods.
Both the Green's function Monte Carlo (GFMC) method and the
auxiliary-field diffusion Monte Carlo (AFDMC) method rely on projection
in imaginary time $\tau$, \begin{equation}
\lim_{\tau\rightarrow\infty}e^{-H\tau}\ket{\Psi_T}
\rightarrow\ket{\Psi_0}, \end{equation} with $H$ the Hamiltonian of the
system and $\ket{\Psi_T}$ a trial wave function not orthogonal to the
many-body ground state $\ket{\Psi_0}$.
For a recent review of developments and applications of QMC methods in
nuclear physics, see Ref.~\cite{carlson2014}.
Recently, we have developed local chiral EFT interactions for use with
QMC methods~\cite{gezerlis2013,lynn2014,gezerlis2014,tews2015}, thereby
producing nonperturbative results for testing the chiral expansion
scheme~\cite{lynn2014} and benchmarks for neutron matter up to high
density~\cite{gezerlis2013,gezerlis2014}.
However, these studies were limited to two-nucleon (\nn{}) interactions
only or to an exploratory study of neutron matter with only the
long-range parts of the \nnn{} interaction.

In this Letter, we include consistent \nnn{} interactions at \ntwolo{}
in coordinate space~\cite{tews2015} in GFMC calculations of light nuclei
and \nalpha{} scattering, and in AFDMC calculations of neutron matter.
We fit the two couplings $c_D$ and $c_E$ to the \isotope[4]{He} binding
energy and low-energy \nalpha{} scattering $P$-wave phase shifts.
The latter system has been studied using various approaches; see, for
example, Refs.~\cite{nollett2007,hagen2006,quaglioni2008}.
These observables are expected to be less correlated than fits to
structure properties of $A=3,4$ systems because the spin-orbit and
$T=\tfrac{3}{2}$ components of the \nnn{} interaction enter directly.

In phenomenological \nnn{} models, any short-range parts which arise
from the Fourier transformation of pion exchanges are typically absorbed
into other short-distance structures: We retain these explicitly.
We choose the \nnn{} cutoff $R_\text{\nnn{}}=R_0$, where $R_0$ is the
\nn{} cutoff, and vary the cutoff in the range
$R_0=1.0-1.2$~fm~\cite{gezerlis2013,lynn2014,gezerlis2014,tews2015}.
Note that with a finite cutoff certain ambiguities appear, including the
specific operator form associated with the shorter-range interactions.
In the Fourier transformation of $V_D$, two possible operator structures
arise:
%\begin{widetext}
\begin{subequations}
\begin{align}
\label{eq:vdcoord1}
V_{D1}&=
\frac{g_A c_D m_\pi^2}{96 \pi \Lambda_\chi F_\pi^4}
\sum_{i<j<k}\sum_\text{cyc}\tdott{i}{k}\left[
\vphantom{\frac{8\pi}{m_\pi^2}}
X_{ik}(\vec{r}_{kj})\drtn{\vec{r}_{ij}}\right.\notag\\
&\left.+X_{ik}(\vec{r}_{ij})\drtn{\vec{r}_{kj}}
-\frac{8\pi}{m_\pi^2}\sdots{i}{k}
\drtn{\vec{r}_{ij}}\drtn{\vec{r}_{kj}}\right],\\
&\vphantom{\frac{4\pi}{m_\pi^2}}\notag\\
\label{eq:vdcoord2}
V_{D2}&=
\frac{g_A c_D m_\pi^2}{96 \pi \Lambda_\chi F_\pi^4}
\sum_{i<j<k}\sum_\text{cyc}\tdott{i}{k}\left[
\vphantom{\frac{4\pi}{m_\pi^2}}X_{ik}(\vec{r}_{ik})\right.\notag\\
&\left.-\frac{4\pi}{m_\pi^2}\sdots{i}{k}
\drtn{\vec{r}_{ik}}\right]
\left[\drtn{\vec{r}_{ij}}+\drtn{\vec{r}_{kj}}\right],
\end{align}
\end{subequations}
%\end{widetext}
where $X_{ik}(\vec{r})=[S_{ik}(\vec{r})\,T(r)+
\sdots{i}{k}]Y(r)$ is the coordinate-space pion
propagator, $S_{ik}(\vec{r})=3\gvec{\sigma}_i\!\cdot\!\widehat{\vec{r}}
\gvec{\sigma}_k\!\cdot\!\widehat{\vec{r}}-
\sdots{i}{k}$ is the tensor operator, and the
tensor and Yukawa functions are defined as
$T(r)=1+3/(m_\pi r)+3/(m_\pi r)^2$ and $Y(r)=e^{-m_\pi r}/r$.
The smeared-out delta function $\drtn{r}=
\tfrac{1}{\pi\Gamma(3/4)R_\text{\nnn{}}^3}
e^{-(r/R_\text{\nnn{}})^4}$ and the long-range regulator multiplying
$Y$, $f_\text{long}(r)=1-e^{-(r/R_\text{\nnn{}})^4}$ are consistent
with the choices made in the \nn{}
interaction~\cite{gezerlis2013,lynn2014,gezerlis2014,tews2015}.
The sum $i<j<k$ runs over all particles 1 to $A$, and the cyclic sum
runs over the cyclic permutations of a given triple.

The two possible $V_D$ structures agree in the limit of
$R_\text{\nnn{}}\rightarrow 0$, because the delta functions then enforce
$i=j$ ($k=j$) in the first (second) term, in which
case~\cref{eq:vdcoord1,eq:vdcoord2} would coincide.
The $V_D$ interaction does not distinguish which of the two nucleons in
the contact participates in the pion exchange.
The second choice, $V_{D2}$, can be obtained with the exchange of a
fictitious heavy scalar particle between the two nucleons in the
contact.
This ambiguity was also pointed out in~\cite{navratil2007}.
The differences between~\cref{eq:vdcoord1,eq:vdcoord2} are regulator
effects and therefore higher order in the chiral expansion, but it is
important to investigate how they affect different observables at this
order.

Similar effects arise in the \nnn{} contact interaction $V_E$.
Here, the main ambiguity is the choice of the \nnn{} contact operator.
The same Fierz-rearrangement freedom that allows for a selection of
(mostly) local contact operators in the \nn{} sector up to \ntwolo{}
exists in the \nnn{} sector at this order.
Symmetry considerations allow the choice of one of the following six
operators~\cite{epelbaum2002}:
\begin{equation}
\begin{split}
&\{\mathbbm{1},\sdots{i}{j},\tdott{i}{j},\sdots{i}{j}\tdott{i}{j},\\
\hfill&\sdots{i}{j}\tdott{i}{k},
[(\gvec{\sigma}_i\times\gvec{\sigma}_j)\!\cdot\!\gvec{\sigma}_k]
[(\gvec{\tau}_i\times\gvec{\tau}_j)\!\cdot\!\gvec{\tau}_k]\}.
\end{split}
\label{v3op}
\end{equation}
The usual choice is $\tdott{i}{j}$.
Here, we investigate two other choices: first the operator
$\mathbbm{1}$, and second a projector operator $\mathcal{P}$ on to
triples with $S~\!\!=~\!\!\tfrac{1}{2}$~and~$T=\tfrac{1}{2}$:
\begin{equation}
\mathcal{P}=\frac{1}{36}\Bigl(3-\sum_{i<j}\sdots{i}{j}\Bigr)
\Bigl(3-\sum_{k<\vphantom{j}l}\tdott{k}{l}\Bigr),
\end{equation}
where the sums are over pairs in a given triple.
In the infinite-momentum cutoff limit, only these $S=\tfrac{1}{2},
T=\tfrac{1}{2}$ triples would contribute to $V_E$ due to the Pauli
principle.
Thus, in the following we will explore three possible structures:
\begin{subequations}
\begin{align}
V_{E\tau}&=\frac{c_E}{\Lambda_\chi F_\pi^4}\sum_{i<j<k}\sum_\text{cyc}
\tdott{i}{k}\drtn{\vec{r}_{kj}}\drtn{\vec{r}_{ij}},\\
V_{E\mathbbm{1}}&=\frac{c_E}{\Lambda_\chi
F_\pi^4}\sum_{i<j<k}\sum_\text{cyc}
\drtn{\vec{r}_{kj}}
\drtn{\vec{r}_{ij}},\\
V_{E\mathcal{P}}&=\frac{c_E}{\Lambda_\chi
F_\pi^4}\sum_{i<j<k}\sum_\text{cyc}\mathcal{P}\,
\drtn{\vec{r}_{kj}}
\drtn{\vec{r}_{ij}}.
\end{align}
\end{subequations}
We stress that there exist other possible operator-structure
possibilities for $V_D$ and $V_E$, which will be investigated in future
work.

\begin{figure*}[t]
   \includegraphics[width=\columnwidth]{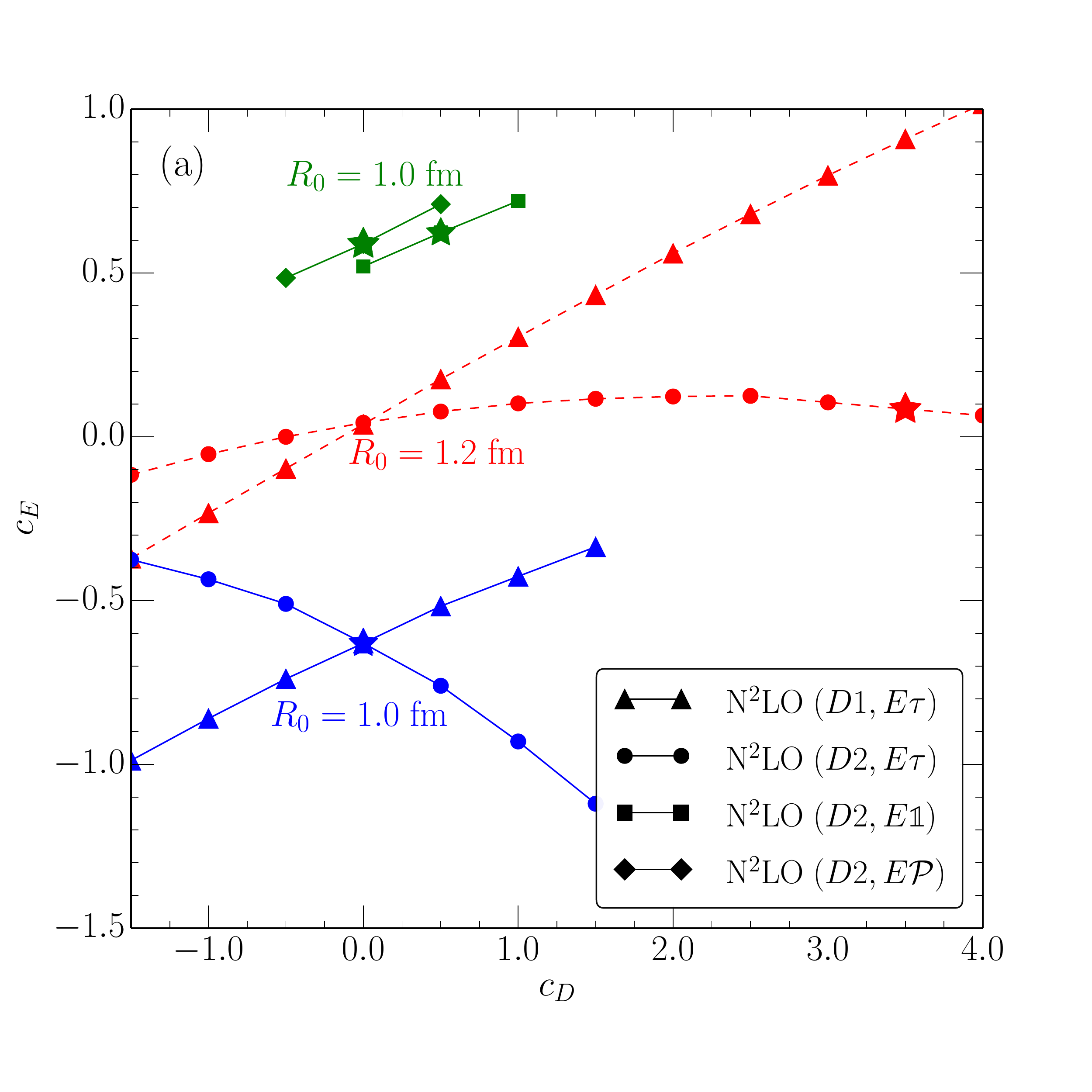}
   \hfill
   \includegraphics[width=\columnwidth]{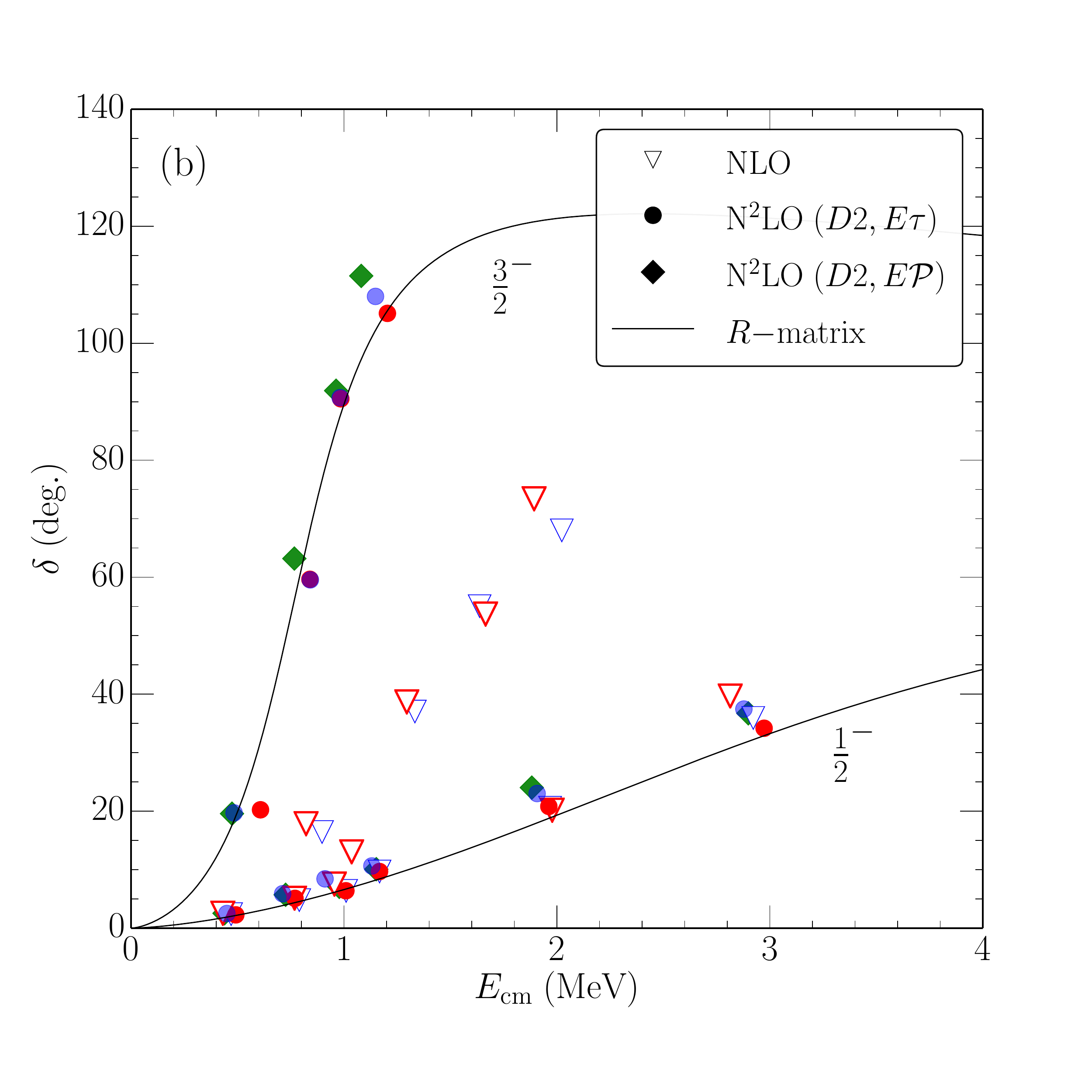}
   \caption{\label{fig:fits}
   (a)~Couplings $c_E$ vs\! $c_D$ obtained by fitting the
   \isotope[4]{He} binding energy for different \nnn{}-operator forms.
   Triangles are obtained by using $V_{D1}$ and $V_{E\tau}$, while the
   other symbols are obtained for $V_{D2}$ and three different 
   $V_E$-operator structures.
   The blue and green lines (lower and upper) correspond to
   $R_0=1.0$~fm,
   while the red lines (central) correspond to $R_0=1.2$~fm.
   The GFMC statistical errors are smaller than the symbols.
   The stars correspond to the values of $c_D$ and $c_E$ 
   which simultaneously fit the \nalpha{} $P$-wave phase shifts 
   (see \cref{tab:cecd} and the right panel).
   No fit to both observables can be obtained for the case with
   $R_0=1.2$~fm and $V_{D1}$.
   (b)~$P$-wave \nalpha{} elastic scattering phase shifts compared with
   an $R$-matrix analysis of experimental data.
   Colors and symbols correspond to the left panel.
   We also include phase shifts calculated at NLO which clearly
   indicate the necessity of \nnn{} interactions to fit the $P$-wave
   splitting.}
\end{figure*}

Having specified all \nnn{} structures, we vary the values of the
couplings $c_D$ and $c_E$ to fit the \isotope[4]{He} binding energy as
shown in~\cref{fig:fits}(a).
We display curves for $V_{D1}$ and $V_{D2}$ using $V_{E\tau}$ and both
cutoffs $R_0=1.0$~fm and $R_0=1.2$~fm.
In addition, we show curves for $V_{D2}$ using the other two possible
$V_E$ structures and the cutoff $R_0=1.0$~fm.
For all of these possibilities, the stars give the values for the
couplings which also fit $P$-wave \nalpha{} scattering phase shifts, as
shown in \cref{fig:fits}(b).
The resulting couplings $c_D$ and $c_E$ are given in~\cref{tab:cecd}.
In all cases $\ev{V_E}$ is repulsive in $\isotope[4]{He}$, except
for the case with $(D2,E\tau)$ with the softer cutoff ($R_0=1.2$~fm),
where it is mildly attractive.

For $R_0=1.0$~fm and $V_{E\tau}$, $c_D~\approx~0$ and both forms of
$V_D$ simultaneously fit the \isotope[4]{He} binding energy and the
$P$-wave \nalpha{} scattering phase shifts [see~\cref{fig:fits}(b)].
However, in the softer-cutoff case $R_0=1.2$~fm, $V_{D1}$ and $V_{D2}$
lead to different couplings.
For $V_{D1}$, the splitting between the two $P$ waves appears to
saturate in $c_D$ for values of $c_D>2$; e.g., the $\tfrac{3}{2}^-$
phase shift for $c_D=2.0,\ 3.0,\ \text{and}\ 5.0$ at $E_\text{cm}=1.3\
\text{MeV}$ are each $\sim75\ \text{deg}$, which is $\sim35\
\text{deg}$ below the $R$-matrix value.
Since we cannot fit the $P$-wave \nalpha{} scattering phase shifts in
this case ($V_{D1}$ and $R_0=1.2$~fm), we do not consider it in the
following.
Instead, for $V_{D2}$ and $R_0=1.2$~fm, the splitting can be fit, as is
evident from~\cref{fig:fits}(b).
For $V_{D2}$ using $V_{E\mathbbm{1}}$ or $V_{E\mathcal{P}}$ and
$R_0=1.0$~fm, both the \isotope[4]{He} binding energy and the $P$-wave
\nalpha{} scattering phase shifts can be simultaneously fit: We show
only the case with $V_{E\mathcal{P}}$ in~\cref{fig:fits}(b).
There, we also show the next-to-leading order (NLO) results which are a
clear indication that
\nnn{} forces are necessary to properly describe \nalpha{} scattering.
Similar results have been found in
Refs.~\cite{hupin2013,hupin2014,quaglioni2015}.
Because $A=3,4$ systems (futher discussed below) are largely insensitive
to odd-parity partial
waves, we find no significant dependence on the choice of structures in
$V_D$.
However, our results in \nalpha{} $P$-wave scattering show a substantial
sensitivity: $V_{D1}$ appears to have a smaller effect than $V_{D2}$.

\begin{figure}[hb!]
   \begin{center}
   \includegraphics[width=1.08\columnwidth]{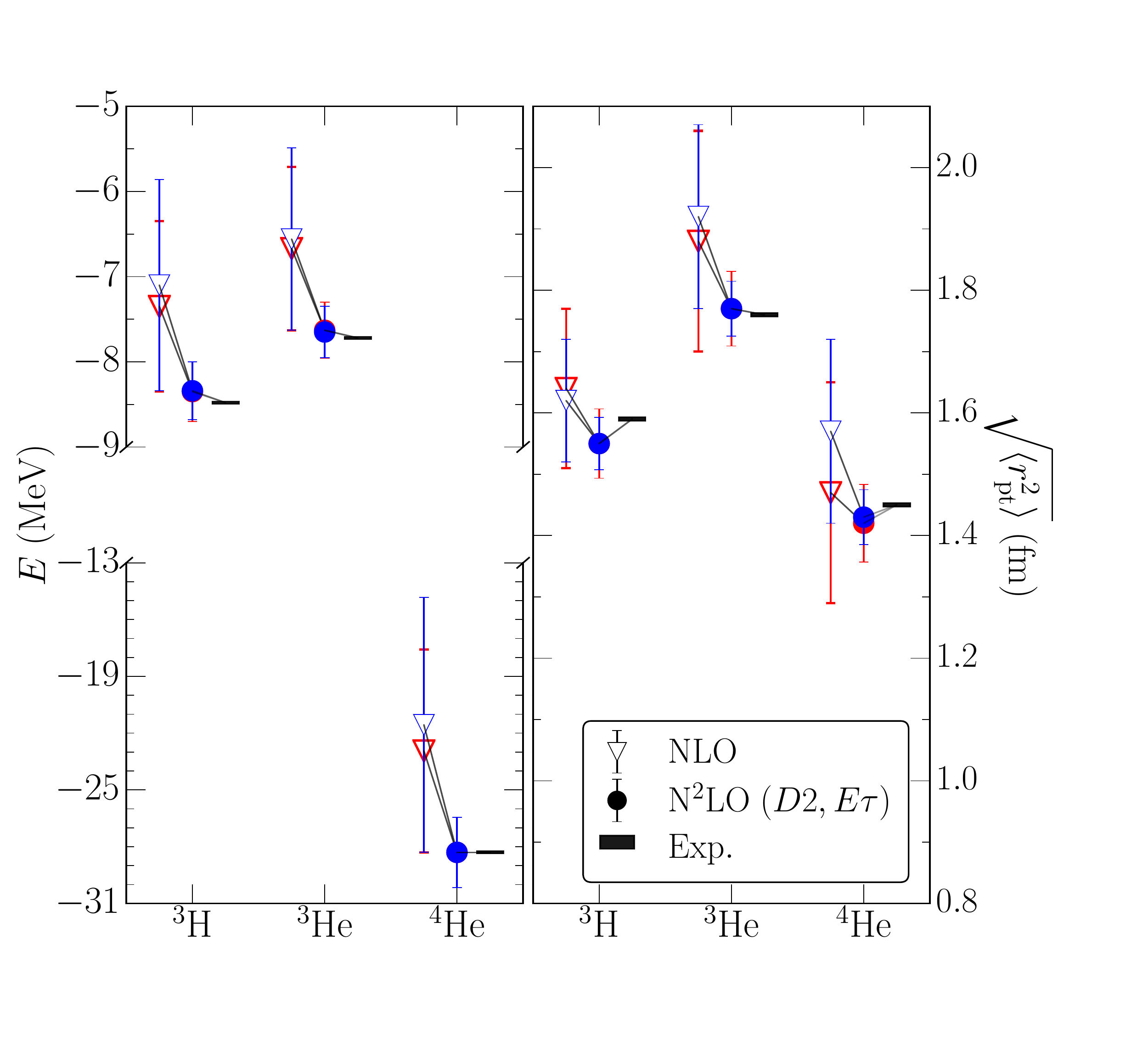}
   \caption{\label{fig:lightnuclei}
   Ground-state energies and point proton radii for $A=3,4$ nuclei
   calculated at NLO and N$^2$LO (with $V_{D2}$ and $V_{E\tau}$)
   compared with experiment.
   Blue (red) symbols correspond to $R_0=1.0$~fm ($R_0=1.2$~fm).
   The errors are obtained as described in the text and also include the
   GFMC statistical uncertainties.}
   \end{center}
\end{figure}

\begin{figure}[t!]
   \includegraphics[width=\columnwidth]{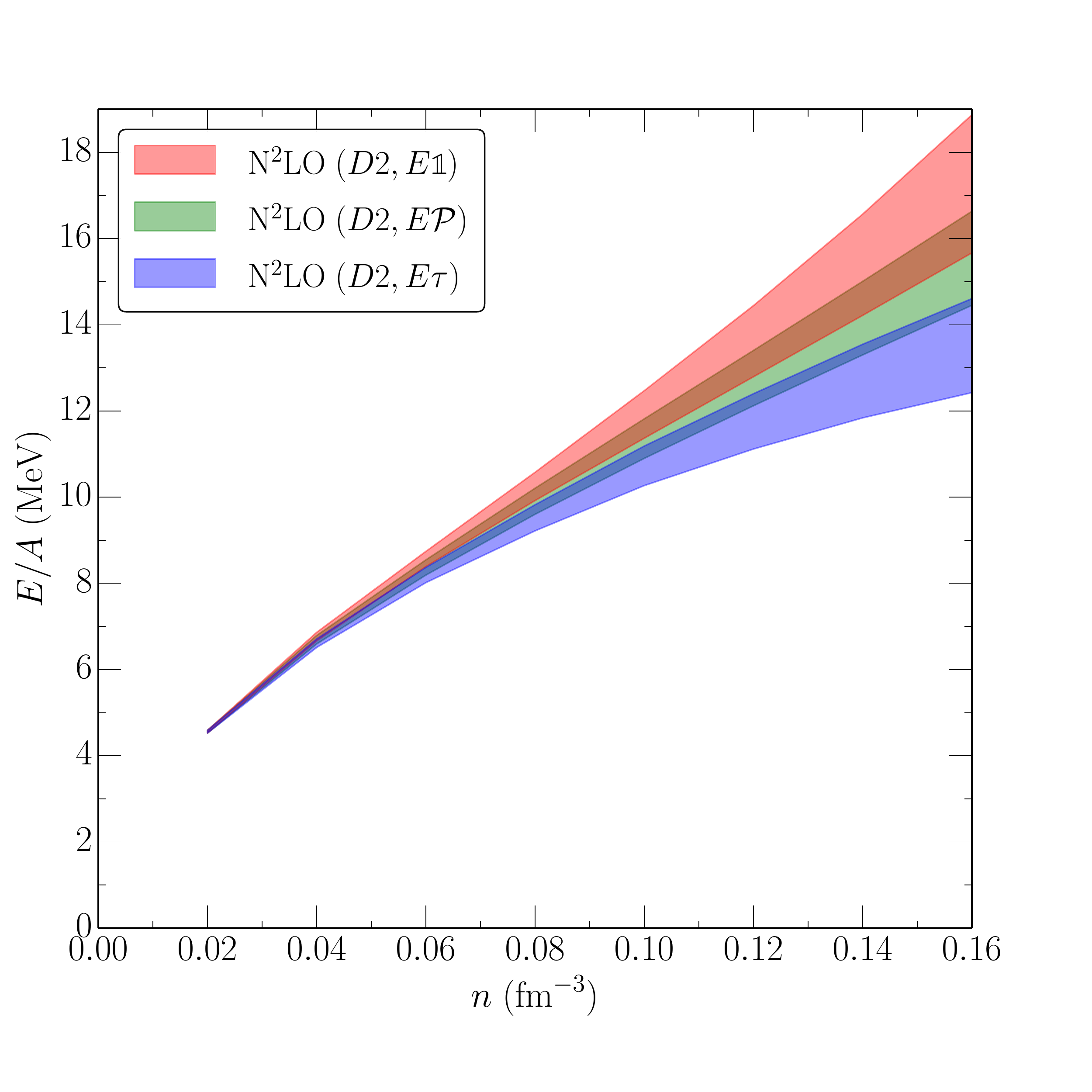}
   \caption{\label{fig:ekmeosnm}
   The energy per particle in neutron matter as a function of density
   for the \nn{} and full \nnn{} interactions at N$^2$LO with
   $R_0=1.0$~fm.
   We use $V_{D2}$ and different \nnn{} contact structures: The blue
   band corresponds to $V_{E\tau}$, the red band to $V_{E\mathbbm{1}}$,
   and the green band to $V_{E\mathcal{P}}$.
   The green band coincides with the $\text{\nn{}}+2\pi$-exchange-only
   result because both $V_D$ and $V_E$ vanish in this case.
   The bands are calculated as described in the text.}
\end{figure}

In~\cref{fig:lightnuclei}, we show ground-state energies and
point proton radii for $A=3,4$ nuclei at NLO and N$^2$LO using $V_{D2}$
and $V_{E\tau}$ for $R_0~\!\!=~\!\!1.0$~fm and $R_0~\!\!=~\!\!1.2$~fm, in
comparison with experiment.
The ground-state energies of the $A=3$ systems compare well with
experimental values.
The ground-state energy of \isotope[4]{He} is used in fitting $c_D$ and
$c_E$, and so it is forced to match the experimental value to within
$\approx 0.03$ MeV.
\begin{table}
\caption{\label{tab:cecd}Fit values for the couplings $c_D$ and $c_E$
for different choices of \nnn{} forces and cutoffs.}
{\renewcommand{\arraystretch}{1.30}
\begin{ruledtabular}
\begin{tabular}{ccdd}
$V_\text{\nnn{}}$&$R_0$ (fm)&\multicolumn{1}{c}{$c_E$}&
\multicolumn{1}{c}{$c_D$}\\
\hline
\multirow{2}{*}{\ntwolo{} $(D1,E\tau)$}&$1.0$&-0.63&0.0\\
&1.2&&\\
\multirow{2}{*}{\ntwolo{} $(D2,E\tau)$}&$1.0$&-0.63&0.0\\
&$1.2$&0.09&3.5\\
{\ntwolo{} $(D2,E\mathbbm{1})$}&$1.0$&0.62&0.5\\
{\ntwolo{} $(D2,E\mathcal{P})$}&$1.0$&0.59&0.0\\
\end{tabular}
\end{ruledtabular}}
\end{table}
The point proton radii also compare well with values extracted from
experiment.
The theoretical uncertainty at each order is estimated through the
expected size of higher-order contributions; see Ref.~\cite{EKM} for
details.
We include results from LO, NLO, and N$^2$LO in the analysis using the
Fermi momentum and the pion mass as the small scales for neutron matter
(discussed below) and nuclei, respectively.
The error bars presented here are comparable to those shown in
Ref.~\cite{binder2015}, although it is worth emphasizing that our
calculations represent a complete estimate of the uncertainty at
\ntwolo{} since we include \nnn{} interactions.
Other choices for \nnn{} structures give similar results.

It is noteworthy that \nn{} and \nnn{} interactions derived from chiral
EFT up to \ntwolo{} have sufficient freedom such that \nalpha{}
scattering phase shifts in~\cref{fig:fits}(b) and properties of light
nuclei in~\cref{fig:lightnuclei} can be simultaneously described.
The failures of the Urbana IX model in underbinding nuclei and
underpredicting the spin-orbit splitting in neutron-rich systems,
including the \nalpha{}, system were among the factors motivating the
addition of the three-pion exchange diagrams in the Illinois \nnn{}
models~\cite{pieper2001.2}.
Our results show that chiral \nnn{} forces at \ntwolo{}, including the
shorter-range parts in the pion exchanges, allow the simultaneous fit.
These interactions should be tested further in light $p$-shell nuclei.

Finally, we study the full chiral N$^2$LO forces, including all \nnn{}
contributions, in neutron matter to extend the results from
Ref.~\cite{tews2015}.
More specifically, we examine the effects of different $V_D$ and $V_E$
structures on the equation of state of neutron matter.
Although these terms vanish in the limit of infinite cutoff, they
contribute for finite cutoffs.
In \cref{fig:ekmeosnm} we show results for the neutron matter energy per
particle as a function of the density calculated with the AFDMC method
described in Refs.~\cite{gandolfi2009,carlson2014}.
We show the energies for $R_0=1.0$~fm for the \nn{} and full \nnn{}
interactions.
We use $V_{D2}$ and the three different $V_E$ structures: $V_{E\tau}$
(blue band), $V_{E\mathbbm{1}}$ (red band), and $V_{E\mathcal{P}}$
(green band).
The error bands are determined as in the light nuclei case.
The $V_{E\mathcal{P}}$ interaction fits $A=4,5$ with a vanishing $c_D$;
hence, this choice of $V_E$ leads to an equation of state identical to
the equation of state with $\text{\nn{}}+V_C$ as in Ref.~\cite{tews2015}
(the projector $\mathcal{P}$ is zero for pure neutron systems),
and qualitatively similar to previous results using chiral interactions
at \ntwolo{}~\cite{hagen2014} and next-to-next-to-next-to-leading
order~\cite{tews2013.2}.

As discussed, the contributions of $V_D$ and $V_E$ are only regulator
effects for neutrons.
However, they are sizable and result in a larger error band.
At saturation density $n_0\sim 0.16\text{ fm}^{-3}$, the difference of
the central value of the energy per neutron after inclusion of the
\nnn{} contacts $V_{E\mathbbm{1}}$ or $V_{E\tau}$ is $\sim2$~MeV,
leading to a total error band with a range of $\sim6.5$~MeV when
considering different $V_E$ structures.
This relatively large uncertainty can be qualitatively explained when
considering the following effects.
Because the expectation value
$\langle\sum_{i<j}\tdott{i}{j}\rangle$ has a sign
opposite to that of the expectation value $\ev{\mathbbm{1}}$ in
\isotope[4]{He}, $c_E$ will also have opposite signs in the two cases to
fit the binding energy.
However, in neutron matter both operators are the same, spreading the
uncertainty band.
A similar argument was made in Ref.~\cite{Lovato:2012}.

With the regulators used here, the Fierz-rearrangement invariance valid
at infinite cutoff is only approximate at finite cutoff, and hence the
different choices of $V_D$ and $V_E$ can lead to different results.
The different local structures can lead to finite relative $P$-wave 
contributions.
These can be eliminated by choosing $V_{E\mathcal{P}}$, which has a
projection onto even-parity waves (predominantly $S$ waves).
The usual nonlocal regulator in momentum space does not couple $S$ and
$P$ waves.

In conclusion, we find for the first time that chiral interactions can
simultaneously fit light nuclei and low-energy $P$-wave \nalpha{}
scattering and provide reasonable estimates for the neutron matter
equation of state.
Other commonly used phenomenological \nnn{} models do not provide this
capability.
These chiral forces should be tested in light $p$-shell nuclei,
medium-mass nuclei, and isospin-symmetric nuclear matter to gauge their
ability to describe global properties of nuclear systems.

We also find that the ambiguities associated with contact-operator
choices can be significant when moving from light nuclei to neutron
matter and possibly to medium-mass nuclei, where the $T=\tfrac{3}{2}$
triples play a more significant role.
The reason for the sizable impact may be the regulators used here, which
break the Fierz-rearrangement invariance, making further investigations
of regulator choices a priority.
The impact of these ambiguities in the contact operators can contribute
to the uncertainties and needs to be studied further.

\acknowledgments{We thank G.~Hale for useful discussions and for
providing us with the $R$-matrix analysis of the \nalpha{} phase shifts.
We also thank A.~Dyhdalo, E.~Epelbaum, R.~J.~Furnstahl, K.~Hebeler, and
A.~Lovato for useful discussions.
This work was supported by the NUCLEI SciDAC program, the U.S. DOE under
Contract No. DE-AC52-06NA25396, ERC Grant No. 307986 STRONGINT, the
Natural Sciences and Engineering Research Council of Canada, the LANL
LDRD program, and the NSF under Grant No. PHY-1404405.
Computational resources have been provided by Los Alamos Open
Supercomputing and the J\"ulich Supercomputing Center.
We also used resources provided by NERSC, which is supported by the U.S.
DOE under Contract No. DE-AC02-05CH11231.

\bibliography{threen_qmc_nuclei}
\end{document}